\documentclass[12pt]{article}
\usepackage{psfrag}
\usepackage{epsfig}     
\begin{document}
\begin{center}
{\LARGE {\bf Breakdown of Heterogeneous Materials}}
\end{center}
\bigskip
\begin{center} {\large Purusattam Ray} \footnote{E-mail: ray@imsc.res.in} \end{center}
\begin{center} 
{\large The Institute of Mathematical Sciences, Taramani, Chennai 600 113, India}
\end{center}

\bigskip
\bigskip
\bigskip
\begin{center} {\large{\bf ABSTRACT}} \end{center}    


\bigskip

\noindent
We discuss the threshold activated extremal dynamics that is
prevalent in the breakdown processes in heterogeneous materials.
We model such systems by an elastic spring network with
random breaking thresholds assigned to the springs.
Results are obtained from molecular dynamics simulation of the
system under constant stress and constant strain conditions.
We find that the distribution $P(m)$ of the avalanches of size $m$, 
caused by the rupturing of the springs till the failure of 
the network, decays as
a power-law: $P(m) \sim m^{-\alpha}$, where $\alpha$ can be
closely approximated to $5/2$. The average avalanche size $<m>$
diverges as $<m> \sim (F_c - F)^{-1/2}$ close to the stress $F_c$
at which the total failure of the network occurs. We study the
time evolution of the breakdown process: we find that the bonds 
rupture randomly over the network at initial times but the 
rupturing becomes highly correlated at late times to give rise 
to a well-defined macroscopic crack.

\bigskip



\bigskip


\newpage

\noindent
It is known for a long time that defects play a crucial role in
the process of nucleation of fracture in a material. A complete 
theoretical analysis of fracture starting from the elasticity
theory and the deformation of elastic field around a defect is, 
however, viable only in very specific cases: like that of
an isolated single defect in the form of a microcrack of suitable
geometrical shape \cite{griffith}.
In a heterogeneous medium, consisting of defects of various
kinds, shapes and vulnerabilities (from the point of nucleation
of fracture) distributed over the medium, fracture phenomena
becomes extremely complex due to the cooperative role played by 
the interacting defects over a wide range of length and
time scales. Since fracture, at any stage, develops from the
most vulnerable defect (weakest link of a chain), a theory based
on a continuous coarse-grained description of fracture is
untenable and a realistic computer simulation is almost unfeasible.

\bigskip

\noindent
Most engineering as well as many natural materials like rocks,
wood, glass (cellular), composite materials (fibre-glass, plaster..)
are examples of heterogeneous systems. These materials, 
though widely different in their physical properties and chemical
composition, show characteristic features prior to fracture when  
they are subjected to increasing stress or strain.
In these materials, fracture does not develop from a single crack
or a microcrack, rather, the macroscopic crack is preceded
by myriads of microscopic crack nucleating from the defects
and the final breakdown results from the birth, growth and coalescence
of these microcracks. The formation of the microcracks are accompanied
by release in the stored elastic energy which come out as acoustic
signals of varying amplitude (energy). These signals are recorded
in experiments and analysed. The experiments show that 
the probability density $N(\epsilon)$ of microfractures
with energy between $\epsilon$ and $\epsilon + d\epsilon$, follows
a power-law: $N(\epsilon) \sim \epsilon^{-\beta}$. Different materials
are characterised by different values of $\beta$: 1.25 in paper
\cite{saliminen}, 1.3 in synthetic plaster \cite{petri}, 1.51 in wood
\cite{garciamartin}, 1.9 in fiberglass \cite{guarino} and
1.5 in cellular glass \cite{maes}.
On the other hand, the cumulative energy emitted while approaching the
fracture also shows power-law in situations where stress is
controlled. For instance, in \cite{garciamartin,guarino}, it was
found that $E(P) \sim [(P_c-P)/P_c]^{-\lambda}$, where $E(P)$ is
the cumulative energy released up to pressure $P$ and $P_c$ is the
critical pressure at which the macroscopic failure occurs. The
exponent $\lambda = 0.27$ seems to be universal for various
substances and loading conditions. Similarly the acoustic emission
$E$ of a material under a constant load (pressure) at time $t$
measured as a function of time shows $E \sim (t_c-t)^{-\lambda}$
where $t_c$ is the time required for the complete failure of the
material \cite{guarino}.

\bigskip

\noindent
The scale-invariance manifested in
the power-law form of the energy density $N(\epsilon)$ and its
temporal correlation tells us
that the developement of fracture in heterogeneous systems does
not take place from a single microcrack like what happens in
Griffith-like nucleation of fracture \cite{lawn}. Rather, fracture 
here is a strongly correlated phenomena where it develops over a
large length and time scales from the growth and coalescence of
microcracks in a self-similar manner. The crucial points here are
that a defect starts to grow only when its stress intensity
exceeds the static fatigue limit of the material (like that in a
Griffith crack). It is only when a critical tension is exceeded
(threshold mechanism) that the self restoration of microdefects
is no longer possible leading to the nucleation of fracture. 
Also at any stress level, only the most vulnerable defect
(weakest link of a chain) grows (extremal mechanism). Fracture
in heterogeneous systems then corresponds to the dynamical response
of a threshold and extremal dynamical system to an
external driving (stress or strain). The system under stress has 
a large number of microscopic metastable states differing in
internal stress distribution and crack structure and the dynamics
takes the system from one metastable state to another by nucleating
a microcrack and emitting energy thereby.

\bigskip

\noindent
Here, we intend to understand the nucleation and the subsequent
propagation of fracture in heterogeneous media from the point of
view of statistical physics. We consider a simplified picture of
the heterogeneous systems and do not take into account the full
details of the defects or their effects on the elastic
response of the system as the defects grow. We consider a discrete 
two-dimensional lattice where the bonds are Hookean springs (of 
identical spring constant) and mimic the heterogeneity by assigning 
a random breaking threshold $\tau$ drawn from a distribution 
$P(\tau)$ to each of the springs. The network is subjected to 
a tensile stress in both the $x$ and $y$-directions.A spring 
behaves like a Hookean except that it can be stretched till the 
threshold value when it ruptures irreversibly. Susequent to a rupture 
the stress is redistributed over the remaining intact part of 
the network. A breaking up of a spring mimics
the nucleation or onset of fracture. It can lead to further
breaking up of the springs and the breaking process
continues or the breaking event may stop whereat the stress
level on the network is to be increased to induce further
breaking. We discuss how the breakdown properties of this
random spring network model give rise to power laws in
breakdown events and compare the results with that of
experimental findings.

\bigskip

\begin{figure}
\vspace{0.2in}
\centerline{
\hbox  {
        \vspace*{0.0cm}
        \epsfxsize=8cm
        \epsfbox{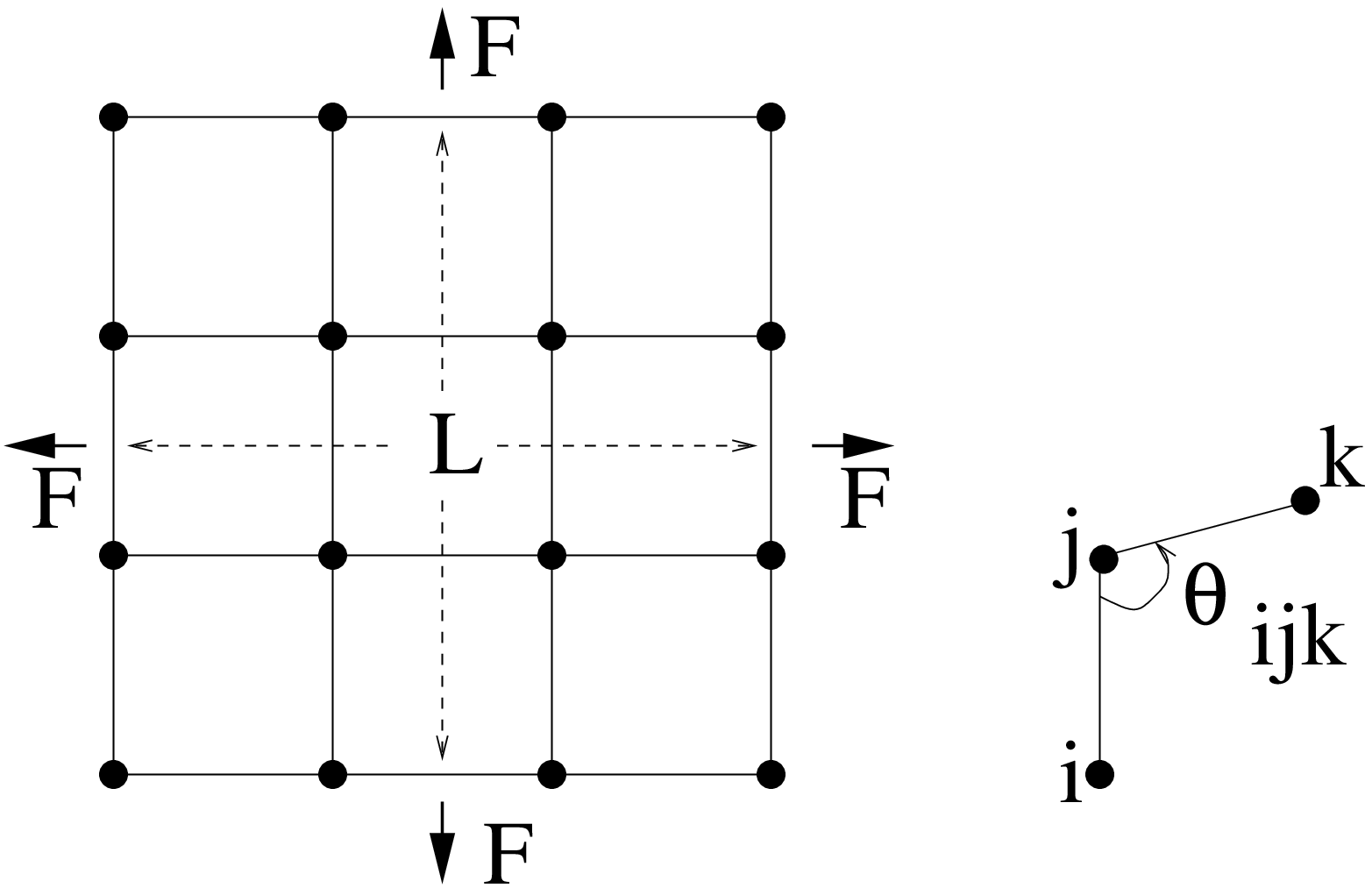}
        \vspace*{0.0cm}
       }
          }
\caption{Schematic diagram of the elastic network used in the simulation.
This is the network prior to the application of the force and with all the 
bonds intact. The bonds are Hookean springs and there is an angular force 
between any two adjacent bonds. The deformations are mesured from the square 
configuration of the network.}
\label{fig7}
\end{figure}

\noindent
The study of fracture in random spring network is carried out
by molecular dynamics simulation. Our system consists of a
$L \times L$ ($L=50$, 100 and 200) square network with central and
rotationally invariant bond-bending forces. The potential
energy of the network is \cite{ray}
$$V=\frac{a}{2}\sum_{<ij>} (\delta r_{ij})^2g_{ij} + \frac{b}{2}\sum_{<ijk>}(\delta\theta_{ijk})^2g_{ij}g_{jk},$$ \nonumber
where $\delta r_{ij}$ is the change in the length of the spring
between the nearest neighbor sites $<ij>$ from its equilibrium
value (which is the lattice spacing in the starting unstretched condition 
and is taken to be unity) and $\delta\theta_{ijk}$ is the
change in the angle between the adjacent springs $ij$ and $jk$
from its equilibrium value which is taken to be $\pi/2$ to ensure the
square lattice structure of the unstretched starting configuration of 
the network (see Fig.~\ref{fig7}). $g_{ij}=1$
if the spring $ij$ is present and $0$ otherwise (when the spring is
broken). $a$ and $b$ are the force constants of the central and
the bond-bending force terms respectively. The dimensionless equation 
of motion
$$\frac{d^2r_i}{dt^2}=\gamma_1\sum_{<j>} (\delta r_{ij})g_{ij} + \gamma_2\sum_{<jk>}(\delta\theta_{ijk})\frac{\partial \theta_{ijk}}{\partial r_i}g_{ij}g_{jk},$$ \nonumber
involves two parameters $\gamma_1=a{t_0}^2/m$ and 
$\gamma_2=bt_0^2/m{l_0}^2$ in terms of the mass $m$ associated 
with the lattice sites, an arbitrary length scale $l_0$ and an 
arbitrary time scale $t_0$. The ratio $\gamma_1/\gamma_2 = a/l_0b$ 
is a characteristic of the system under consideration. This suggests 
that the dynamical features of the network as described by the 
equation of motion do not depend on the choice of the scale of mass 
or time. The obvious choice for $l_0$ is unity which is the lattice 
spacing of the lattice at the unstretched condition. We choose 
$\gamma_1=1.0$ and $\gamma_2=0.1$. The small value of $\gamma_2$, much 
less than the value of $\gamma_1$, allows the fracture to develop without 
much deformation of the network. We start with all the springs intact
so that $g_{ij}=1$ for all neighboring $ij$'s and with each spring
we associate a random breaking threshold $\tau_{ij}$, chosen from a
uniform distribution $P(\tau)\in [0,2]$.

\bigskip

\noindent
We impose a constant external force $F$ on the sites of the boundary
and the system is allowed to evolve dynamically using Verlet's algorithm
\cite{allen},
$$\vec{r_i}(t+\Delta t)=2\vec{r_i}(t)-\vec{r_i}(t-\Delta t)+\vec{F_i}(t)(\Delta t)^2.$$ \nonumber
Here $\vec{F_i}(t)$ is the force (as determined from the potential
energy and boundary condition) and $\vec{r_i}(t)$ is the position 
vector of the site $i$
at time $t$. The simulation involves discrete time $t$ in steps of
$\Delta t$. After $n$ iterations the time elapsed is $n\Delta t$ while 
the real time elapsed is $nt_0\Delta t$. 
To speed up the computation one would wish to choose a large
value of $\Delta t$. However, there is an upper limit to this value
given by the convergence time for the fastest developing components
of the stress distribution, which is generally very small in disordered
systems. We choose $\Delta t=0.01$. Also, we add a small viscous
damping to the evolution to avoid excessive oscillations and to achieve
equilibrium for a given applied force faster. 
For a given applied force, once the system reaches equilibrium, we 
check if any spring $ij$ is stretched beyond its cutoff value 
$\tau_{ij}$ and if this happens the spring is snapped irreversibly 
($g_{ij}$ for that spring is set to zero). Once the springs are broken, 
the system is again brought to equilibrium and the springs are checked 
again to see if the initial set of breaking initiates further rupturing 
of springs. When no more breaking of springs take place the external 
force $F$ is increased in small steps. At each step we compute the 
number of broken bonds, which constitute an avalanche. To average over
disorder, the simulation is repeated for 50 different configurations of
threshold values $\tau$.

\bigskip

\begin{figure}
\vspace{-1in}
\centerline{
\hbox  {
        \vspace*{0.5cm}
        \epsfxsize=6cm
        \epsfbox{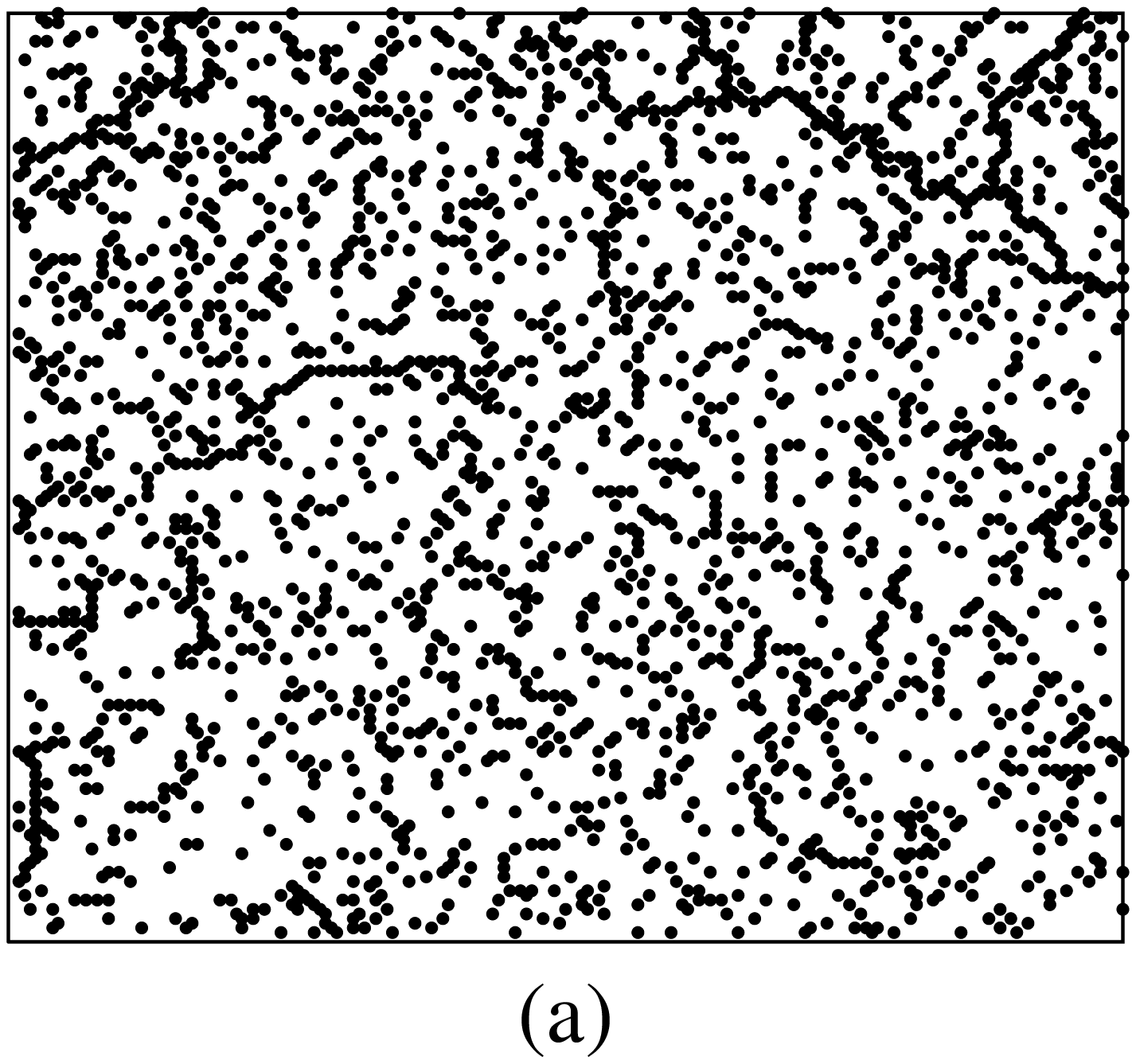}
        \hspace{1.0cm}
        \epsfxsize=6cm
        \epsfbox{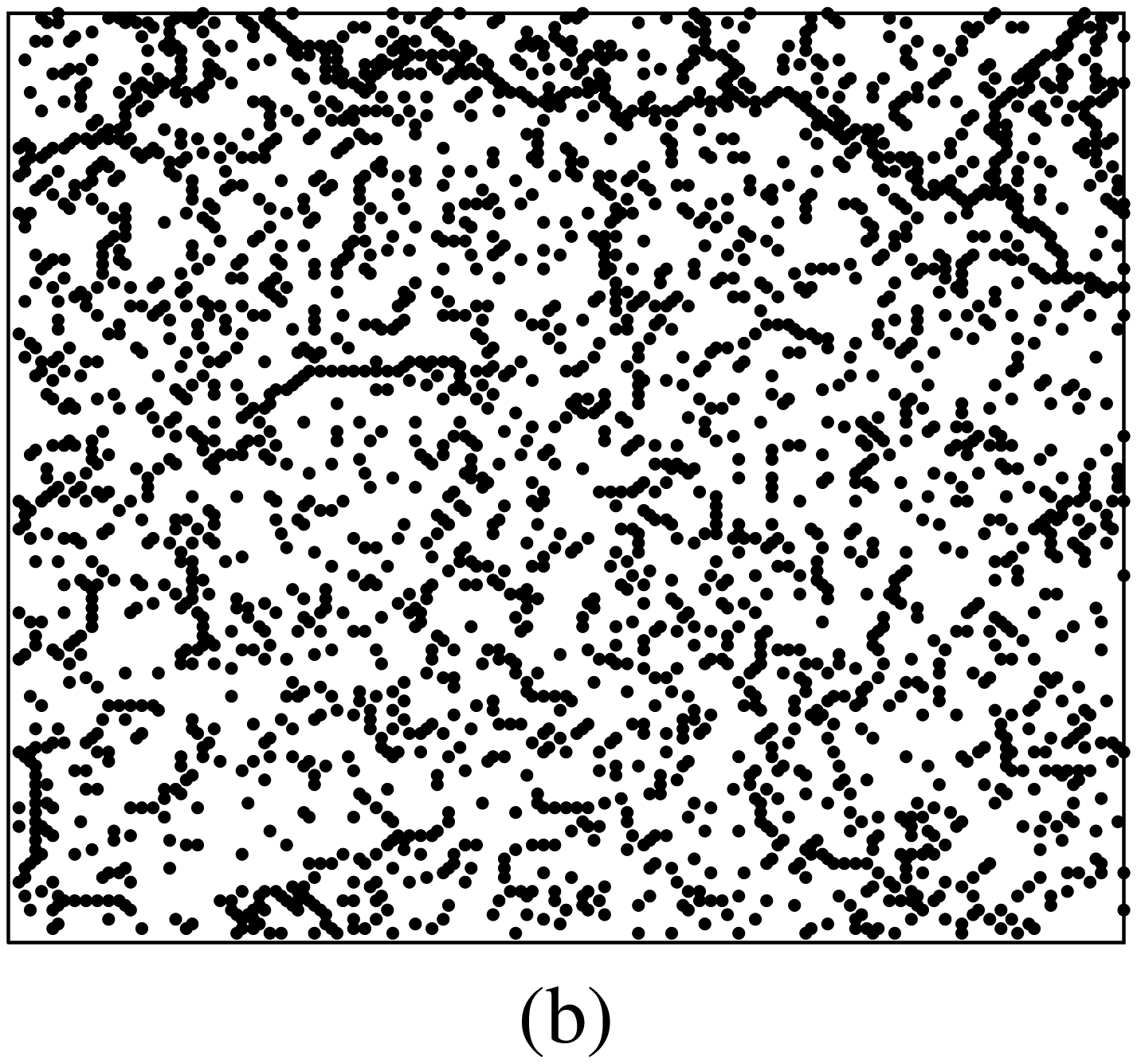}
        \vspace*{0.5cm}
       }
          }
\vspace*{1cm}
\centerline{
\hbox    {
          \vspace*{0.5cm}
          \epsfxsize=6cm
          \epsfbox{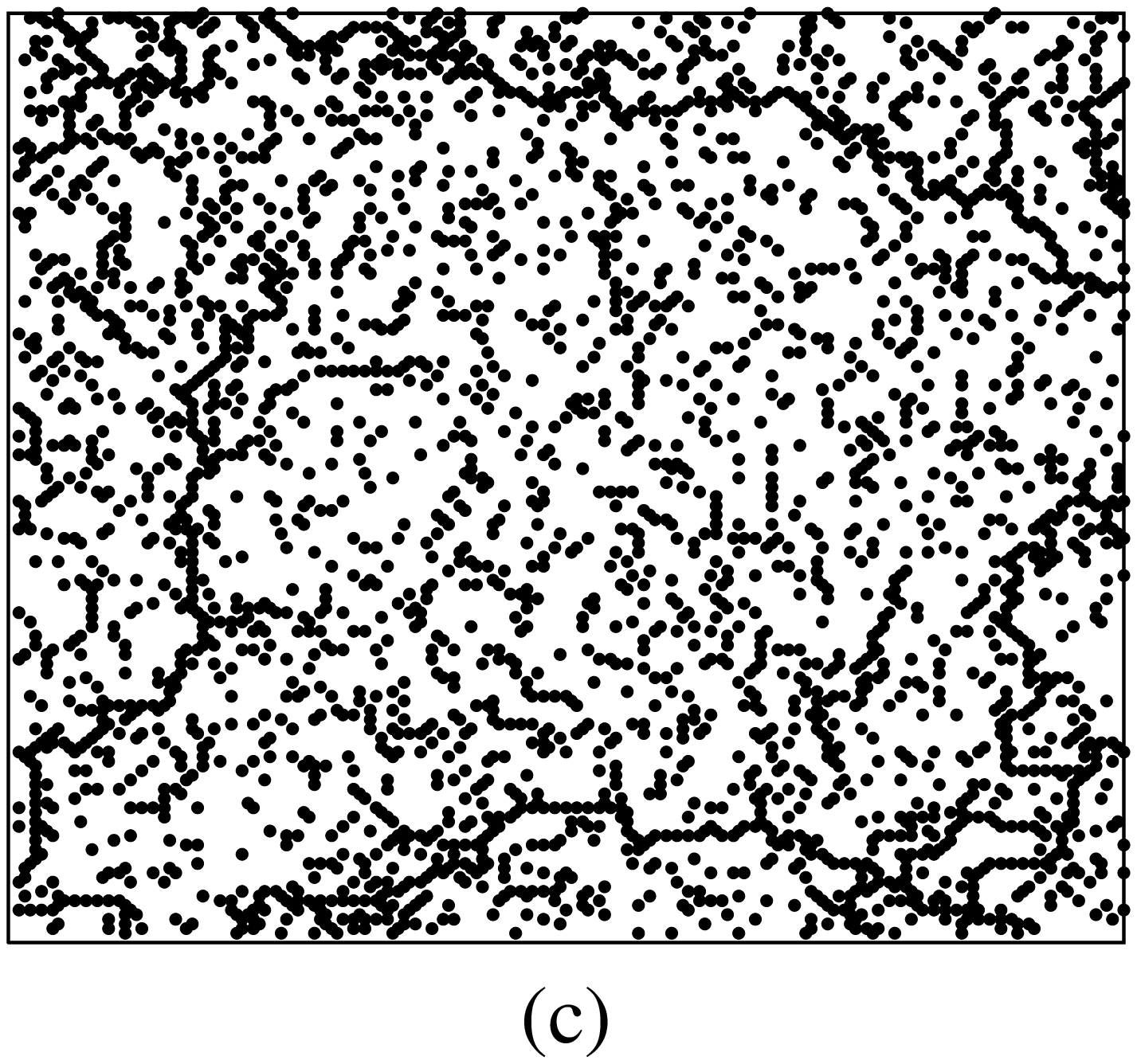}
          \vspace*{0.5cm}
        }
          }     
\caption{Ruptured bonds are shown in black in a $100\times 100$ 
network for stresses (a) $F=0.10$, (b) $F=0.20$ and (c) $F=0.25$.}
\label{fig1}
\end{figure}

\begin{figure}
\vspace{-1in}
\centerline{
\hbox  {
        \vspace*{0.5cm}
        \epsfxsize=6cm
        \epsfbox{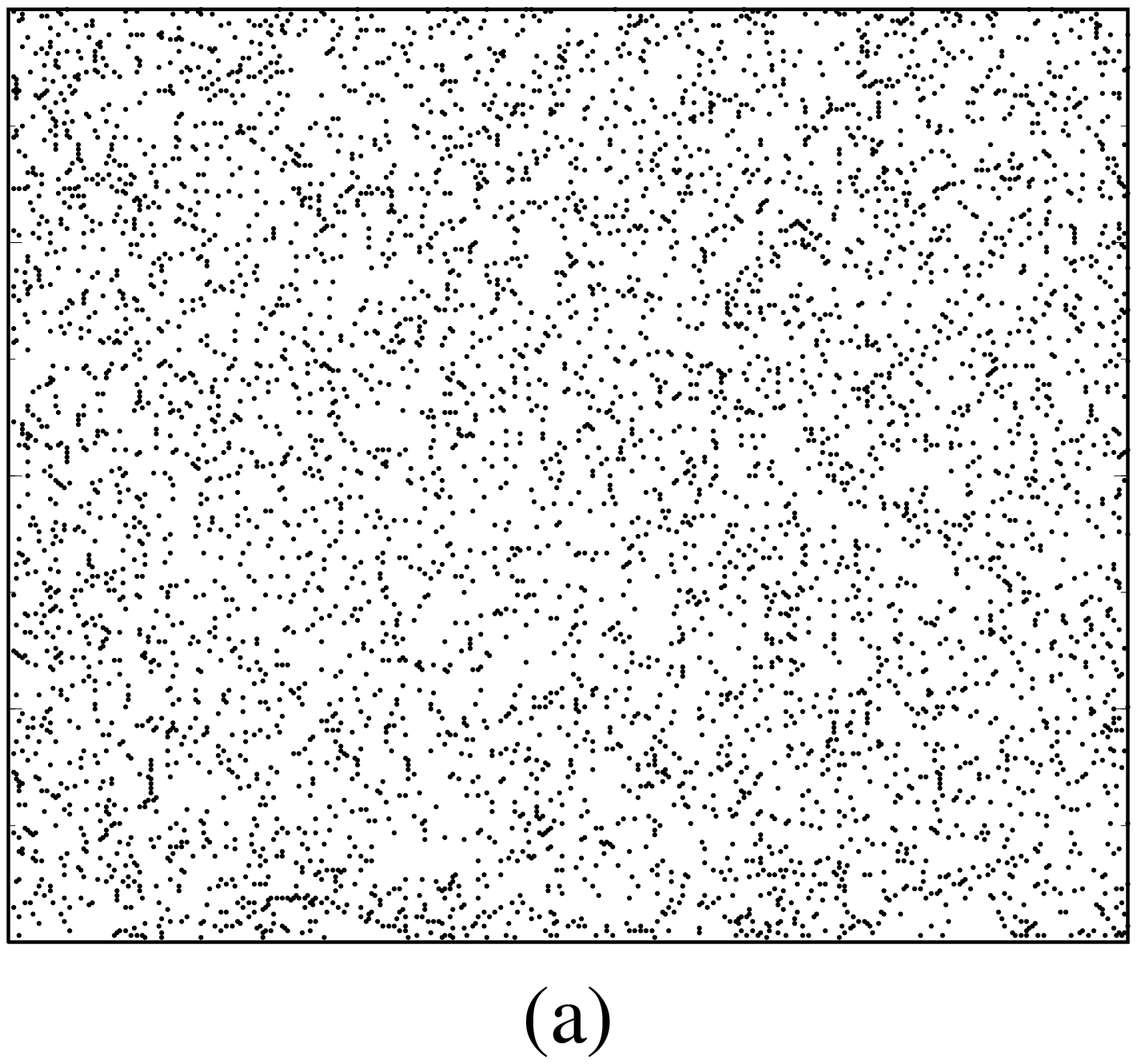}
        \hspace{1.0cm}
        \epsfxsize=6cm
        \epsfbox{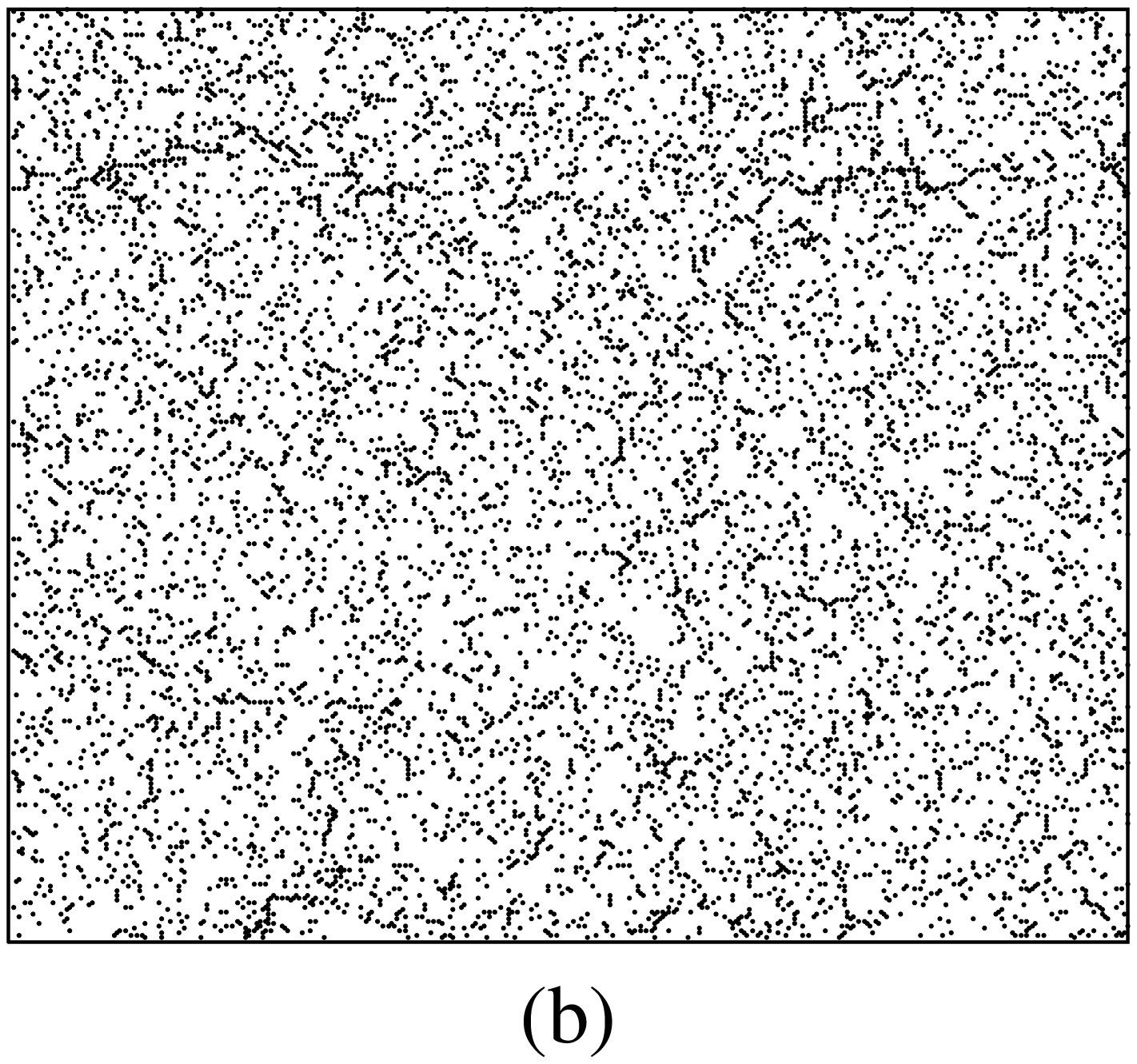}
        \vspace*{0.5cm}
       }
          }
\vspace*{1cm}
\centerline{
\hbox    {
          \vspace*{0.5cm}
          \epsfxsize=6cm
          \epsfbox{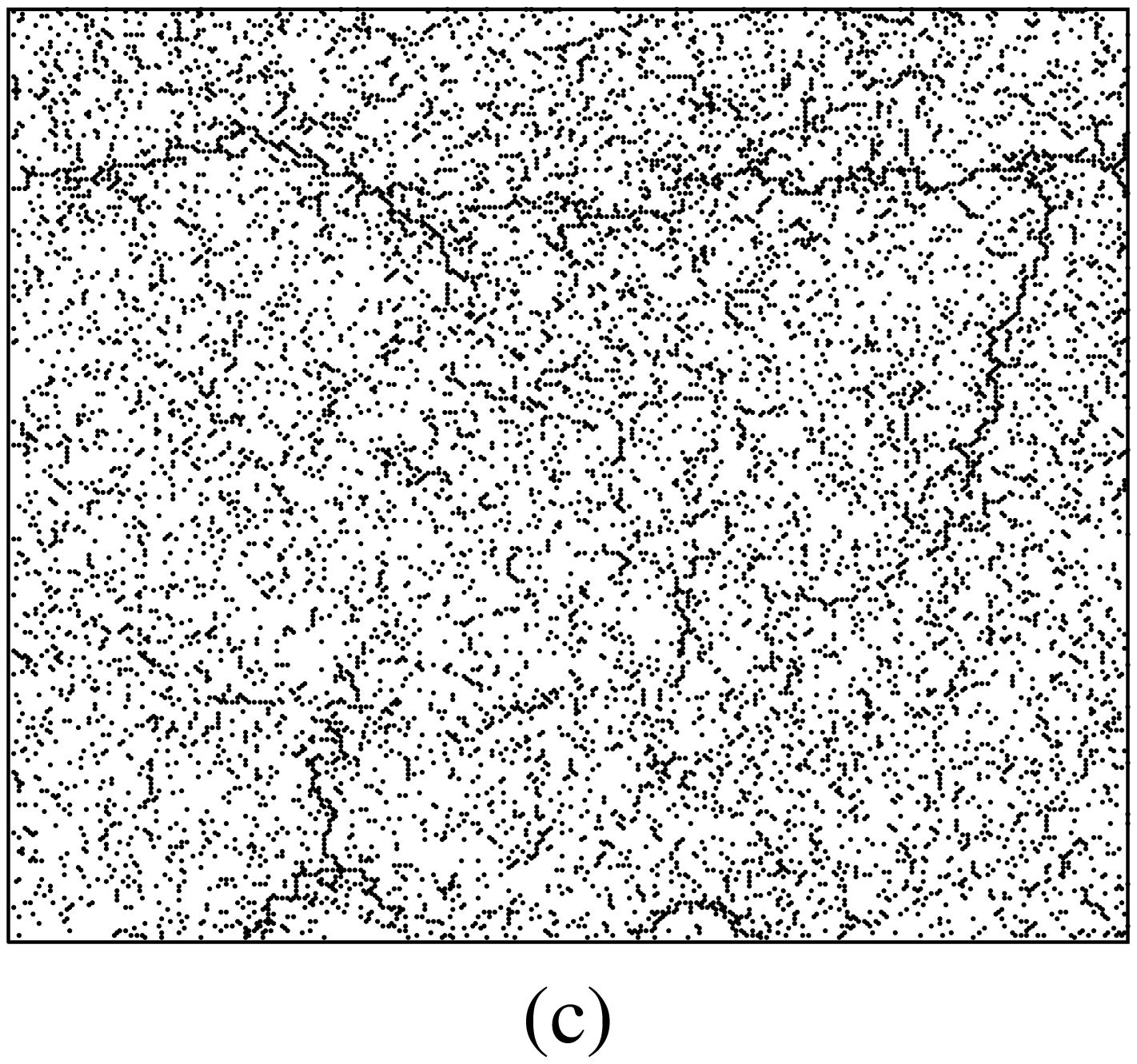}
          \hspace{1.0cm}
          \epsfxsize=6cm
          \epsfbox{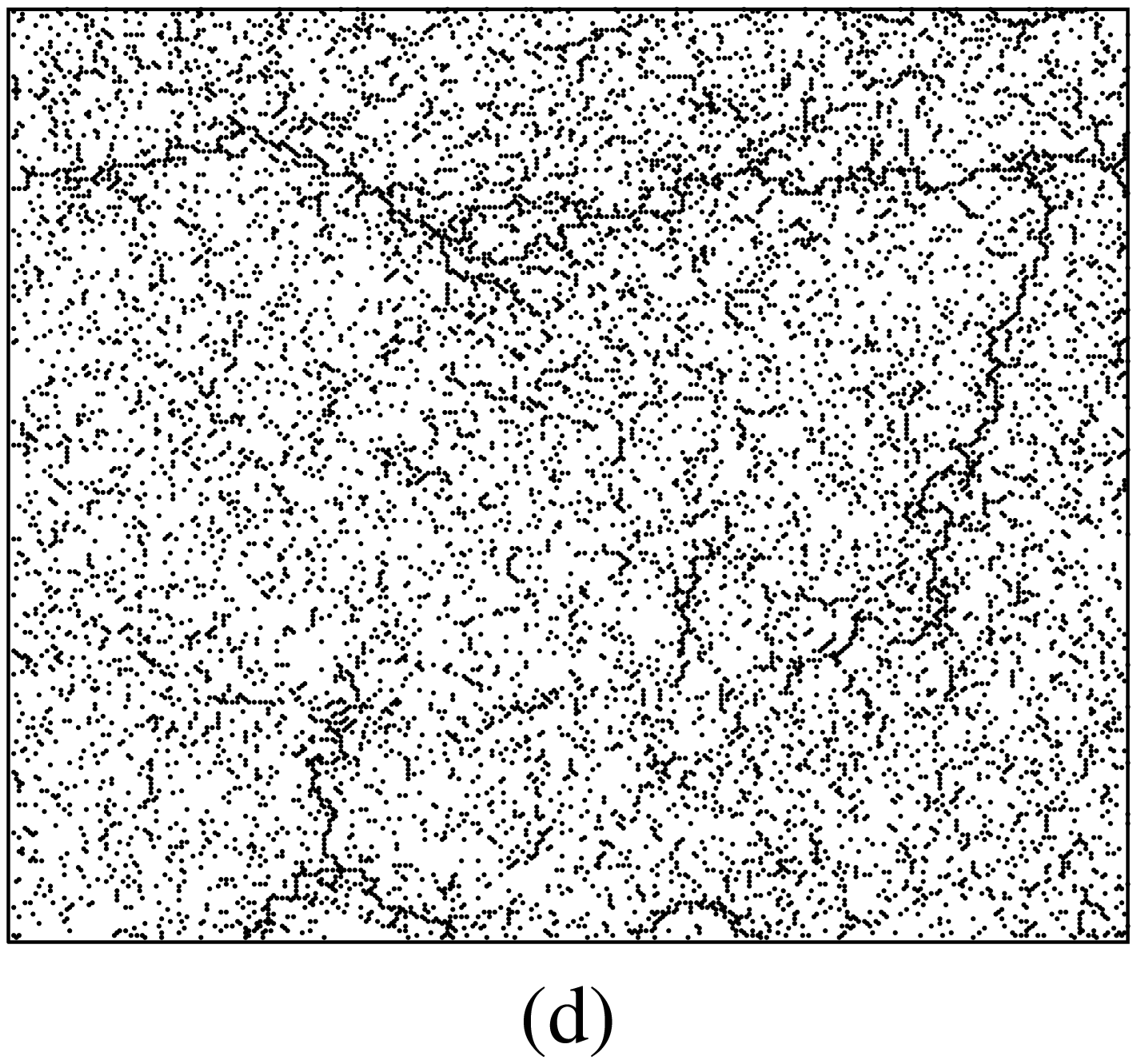}
          \vspace*{0.5cm}
        }
          }     
\caption{Ruptured bonds are shown in black in a $200\times 200$ 
network subjected to the stress $F=0.20$ at molecular dynamics time
steps (a) $t=120000$, (b) $t=240000$, (c) $t=360000$ and $t=480000$.}
\label{fig2}
\end{figure}

\begin{figure} 
\vspace{-1.3in}
\centerline{
\hbox  {
        \vspace*{0.0cm}
        \psfrag{nc1}{$F$}
        \psfrag{nc2}{$n_c$}   
        \epsfxsize=7cm
        \epsfbox{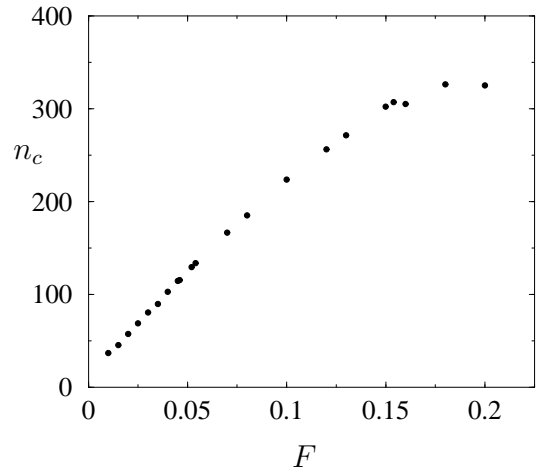}
        \vspace*{0.0cm}
       }
          }
\caption{The number $n_c$ of clusters of ruptured bonds is plotted against 
the stress $F$ in a network of size $L=50\times 50$.}
\label{fig3}
\end{figure}

\begin{figure}
\vspace{0.2in}
\centerline{
\hbox  {
        \vspace*{0.0cm}
        \psfrag{sz1}{$F$}
        \psfrag{sz2}{$s_c$}   
        \epsfxsize=7cm
        \epsfbox{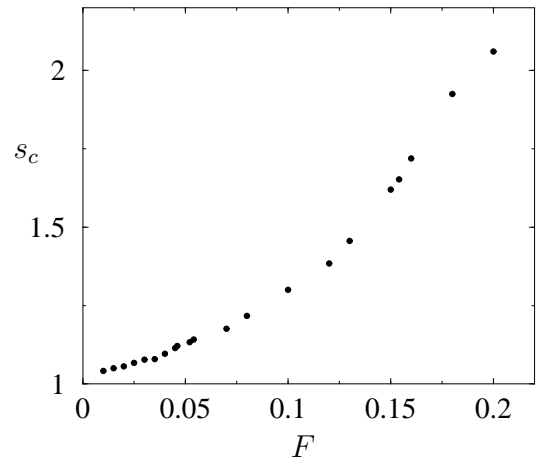}
        \vspace*{0.0cm}
       }
          }
\caption{The average size $s_c$ of the clusters of ruptured bonds 
is plotted against the stress $F$ in a $L=50\times 50$ network.}
\label{fig4}
\end{figure}

\begin{figure}
\vspace{-0.3in}
\centerline{
\hbox  {
        \vspace*{0.5cm}
        \psfrag{m}{$m$}
        \psfrag{P(m)}{$P(m)$}   
        \epsfxsize=7cm
        \epsfbox{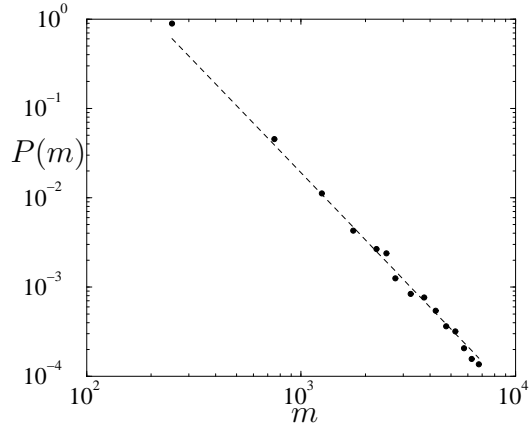}
        \vspace*{0.5cm}
       }
          }
\caption{The size distribution $P(m)$ is plotted against the size 
$m$ of avalanche of ruptured bonds integrated over all the values of 
stress upto the breakdown point $F_c$ in a $L=50\times 50$ network }
\label{fig5}
\end{figure}

\begin{figure}
\centerline{
\hbox  {
        \vspace*{0.5cm}
        \psfrag{form}{$(<m>/L^2)^{-2}$}
        \psfrag{form2}{$F$}   
        \epsfxsize=7cm
        \epsfbox{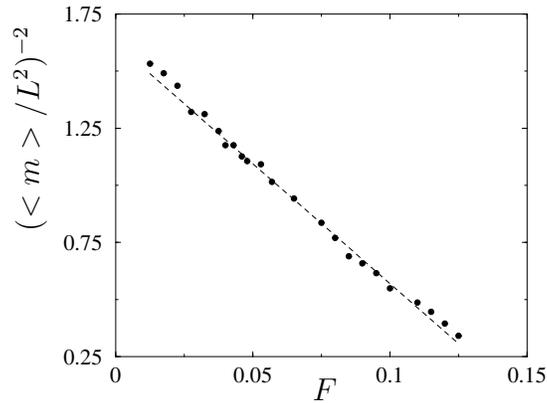}
        \vspace*{0.5cm}
       }
          }
\caption{$(<m>/L^2)^{-2}$ is plotted against $F$, where $<m>$ is the average 
size of the avalanche of ruptured bonds in $L=50\times 50$ network 
integrated upto the stres $F$ 
and $F_c$ is the stress at which the network fails completely.}
\label{fig6}
\end{figure}

\noindent
Our simulation shows that the fracture in our spring network develops
in a series of bursts of spring rupturing processes. In one such
burst, bonds rupture from different parts of the network in a 
random fashion. Fig.~\ref{fig1} shows the ruptured bonds in a
$100\times 100$ lattice for $F =$ 0.10, 0.20 and 0.25.
We find a well defined macroscopic fracture across the network at
$F=0.20$. Below this critical value of $F$ there is no
crack that spans the network and the bonds rupture randomly and 
uniformly over the network. This phenomenon has also been observed 
in the experiment \cite{garciamartin}. Fig.~\ref{fig2} shows the development
of fracture in the network with time. At early times, the bonds
are broken randomly over the network. At later times the
microcracks start coalescing and a large crack develops which
wins over the others and engulf nearby microcracks to form a
crack which spans the system. We keep track of the clusters
formed by the adjacent broken bonds \cite{puru}. In this 
respect, an isolated single 
broken bond form a cluster of size one. The number $n_c$ of such clusters 
grows with the stress $F$ following the relation $n_c=L^2g(F)$ 
(see Fig.~\ref{fig3}), where $g(F)$ is a scaling function of $F$. 
This relation remains valid till the breakdown point indicating that 
the final crack results from sudden coalescence of few large  
microcracks without any drastic change in the number $n_c$. 
The final breakdown resembles a first order transition
and the scaling form of $n_c$ further strengthen this point of view.
Fig.~\ref{fig4} shows the variation of the average size $s_c$ of the 
clusters of ruptured bonds with the stress $F$. We do not find any 
evidence of divergence of $s_c$ which is a strong indication that it 
is not a a second order phase transition. In fact the average 
size $s_c$ remains finite and quite small which suggests that the 
final breakdown is a highly correlated phenomenon involving 
coalescence of very few microcracks. 
Next we consider the distribution $P(m)$ of the size $m$ of burst or 
avalanche (number of bonds that snaps in a burst) integrated 
over all the values of stress $F$ upto the breakdown point $F_c$. 
We find $P(m) \sim m^{-\alpha}$ with $\alpha = 5/2$ as is 
shown in Fig.~\ref{fig5}. This 
amplitude distribution can be transformed into an energy distribution 
(the energy is proportional to the square of the amplitude) giving 
the exponent $\beta = \frac{1+\alpha}{2} = 1.75$. This compares well 
with the experimental 
results. In Fig.~\ref{fig6}, $<m>^{-2}$ is plotted against F and we 
see the linear behavior which suggest $<m> \sim (F_c-F)^{-1/2}$ so that 
the exponent $\lambda=0.5$ in our model.  
   
\bigskip

\noindent 
In conclusion, we see that the dynamical response of a simple  
elastic network in presence of a threshold and extremal dynamical 
rules (assigning a random breaking threshold with each bond and specifying 
the extremal dynamical rule in the bond breaking process)  
produces several features characteristic of fracture in heterogeneous 
materials. It gives the power-law behaviors of the avalanche statistics 
which are observed in the experiments. The simulation shows the right trend 
of development of fracture with time and with stress 
as is observed in experiments. 
It will be interesting to study at what point the stress concentration 
factor comes into play so that starting from random events of bond 
breaking process over the network one ends up with a well-defined 
predominant crack which seems to have the right geometry of a crack 
that we generally find in our day-to-day life. The study of the 
morphology of the crack structure, for example the roughness of the 
crack, is in progress.

\bigskip

\begin{flushleft} {\bf References} \end{flushleft}

\bigskip  

\newpage

\newpage

\bigskip

\vspace*{1in}

\newpage   

\vspace*{1in}

\end{document}